\documentclass{article}

\usepackage{arxiv}

\usepackage[utf8]{inputenc} 
\usepackage[T1]{fontenc}    
\usepackage{hyperref}       
\usepackage{url}            
\usepackage{booktabs}       
\usepackage{amsfonts}       
\usepackage{nicefrac}       
\usepackage{microtype}      
\usepackage{lipsum,amsmath}
\usepackage{graphicx}
\usepackage{natbib}
\setcitestyle{authoryear,open={(},close={)}}

\usepackage{setspace}
\title{MixTwice: large-scale hypothesis testing for peptide arrays by variance mixing}

\author{
  Zihao Zheng \\
  Department of Statistics\\
  Department of Medicine\\
  University of Wisconsin-Madison\\
  Madison, WI, USA\\
  \texttt{zihao.zheng@wisc.edu} \\
  \And
  Aisha M. Mergaert\\
  Department of Medicine\\
  Department of Pathology and Laboratory Medicine\\
  University of Wisconsin-Madison\\
  Madison, WI, USA\\
  \texttt{mergaert@wisc.edu} \\
  \And
  Irene M. Ong\\
  Department of Biostatistics and Medical Informatics\\
  Department of Obstetrics and Gynecology\\
  University of Wisconsin Carbone Comprehensive Cancer Center\\
  University of Wisconsin-Madison\\
  Madison, WI, USA\\
  \texttt{irene.ong@wisc.edu} \\
  \And
  Miriam A. Shelef\\
  Department of Medicine\\
  William S. Middleton Memorial Veterans Hospital\\
  University of Wisconsin-Madison\\
  Madison, WI, USA\\
  \texttt{mshelef@medicine.wisc.edu} \\
  \And
  Michael A. Newton \thanks{To whom the corresponding should be addressed}\\
  Department of Statistics\\
  Department of Biostatistics and Medical Informatics\\
  University of Wisconsin-Madison\\
  Madison, WI, USA\\
  \texttt{newton@biostat.wisc.edu} \\
}

\begin{document}
\maketitle

\begin{abstract}
Peptide microarrays have emerged as a powerful technology in immunoproteomics as they provide a tool to measure the abundance of different antibodies in patient serum samples. The high dimensionality and small sample size of many experiments challenge conventional statistical approaches, including those aiming to control the false discovery rate (FDR). Motivated by limitations in reproducibility and power of current methods, we advance an empirical Bayesian tool that computes local false discovery rate statistics and local false sign rate statistics when provided with data on estimated effects and estimated standard errors from all the measured peptides. As the name suggests,  the \verb+MixTwice+ tool involves the estimation of two mixing distributions, one on underlying effects and one on underlying variance parameters.  Constrained optimization techniques provide for model fitting of mixing distributions under weak shape constraints (unimodality of the effect distribution). Numerical experiments show  that \verb+MixTwice+ can accurately estimate generative parameters and  powerfully identify non-null peptides. In a peptide array study of rheumatoid arthritis (RA), \verb+MixTwice+ recovers meaningful peptide markers  in one case where the signal is weak, and has strong reproducibility properties in one case where the signal is strong. \verb+MixTwice+ is available as an R software package. \href{https://github.com/wiscstatman/MixTwice}{https://github.com/wiscstatman/MixTwice}
\end{abstract}


\section{Introduction}
Peptide microarray technology is used in biology, medicine, and pharmacology to measure various forms of protein interaction. Like other microarrays, a peptide array contains a large number of very small probes arranged on a glass or plastic chip.  Each probe occupies a spatial position on the array, and is comprised of many molecular copies of a short amino-acid sequence (a peptide) anchored to the surface, perhaps 12 to 16 amino acids in length, depending on the design. In antibody profiling experiments, the array is exposed to serum derived from a donor's blood sample; antibodies in the sample that recognize an anchored peptide epitope may bind to the probe. In order to measure these antibody/antigen binding events, a second, fluorescently tagged antibody is applied,  which binds to exposed sites on the already-bound antibodies, providing quantitative readout at probes where there has been sufficient binding of serum antibody recognizing the peptide epitopes. High-density peptide microarrays have emerged as a powerful technology in immunoproteomics, as they enable simultaneous antibody-binding measurements against millions of peptide epitopes. Such arrays have guided the discovery of  markers for viral, bacterial, and parasitic infections \citep{mishra2018diagnosis,tokarz2020identification, bailey2020microarray} and have illuminated the serological response to cancer \citep{yan2019whole} and cancer immunotherapy \citep{hoefges2020thousands}.   The photolithographic design allows for custom arrays, which have benefited studies of autoimmunity, for example, where various forms of post-translational modification (e.g., citrullination)  create targets for autoantibodies~\citep{bailey2017pegivirus,zheng2020disordered}.

The high dimensionality and small sample size of many peptide-array experiments challenge conventional statistical approaches. \cite{zheng2020disordered}, for example, reported a custom peptide-array having 172,828 distinct features and array data from 60 human subjects across several disease subsets. This dimensionality is relatively high compared to gene-expression studies, but quite low compared to other peptide-array studies; arrays that probe the entire human proteome carry over 6 million peptide features, for example.  Methods for large-scale hypothesis testing respond to these challenges, often aiming to control the false discovery rate (FDR) \citep[e.g.,][]{efron2012large}. FDR-controlling procedures are more forgiving than techniques that control the probability of any type I errors (e.g., Bonferroni correction), but they still extract a high penalty for dimensionality in the peptide-array regime involving 10$^5$-10$^6$ features. When additional data are available it may be possible to further limit penalties associated with large-scale testing.

Continuing with \cite{zheng2020disordered}, the authors sought to identify peptides for which antibody binding levels differ between control subjects and rheumatoid arthritis (RA) patients expressing a specific disease marker combination (CCP+ and RF-). Sera from twelve subjects in each group were applied to their custom-built array.  After pre-processing, a univariate  statistic  (t-statistic) measured statistical changes at each peptide. Peptides with the most extreme statistics (and smallest p-values) would be set aside for further validation. In the CCP+RF- RA example, no peptides had a FDR-adjusted p-value less than 10\% by either the Benjamini-Hochberg (BH) method \citep{benjamini1995controlling} or the more sensitive $q-$value method \citep{storey2003positive}, although the latter method estimated that 21\% percent of the peptides in fact have differential binding between the two groups.

Improving power while maintaining robustness and reproducibility is a theme of contemporary large-scale 
inference that we explore in the peptide-array setting. 
The BH and $q-$value procedures yield no discoveries in the CCP+RF- RA example at one conventional FDR level. If this is due to low statistical power, it may not be surprising since these procedures enter quite late in  data analysis, after all p-values have been computed. Procedures that intervene earlier have access to more information, and thereby may have better overall operating characteristics. Efron's local FDR approach, \verb+locFDR+, intervenes on test statistics just prior to p-value computation and has improved power properties in some settings \citep{efron2001empirical}. Independent filtering combines a selection statistic, such as marginal sample variance, and then applies an FDR-controlling procedure to the selected peptides \citep{bourgon2010independent}. Neither \verb+locFDR+ nor independent filtering at 50\% yielded any results in the CCP+RF- RA example, as it happens. We have the same null finding by independent hypothesis weighting (IHW), which generalizes independent filtering in not requiring a specific selection rate \citep{ignatiadis2016data}.

Adaptive Shrinkage (ASH) is a recent innovation for large-scale testing that intervenes after each peptide yields both an estimated effect and an estimated standard error \citep{stephens2017false}.  There are several variations of its empirical Bayesian formulation; when using the $t-$distribution sampling-model version of ASH (say ASH-t), we discover 76 peptides to have differential antibody binding in the CCP+RF- RA comparison, also at 10\% FDR control. This may reflect increased power, and is consistent with numerical studies showing increased power of ASH in many settings.  A recent report from Professor Stephens's group points out a technical limitation of ASH-t that could cause FDR inflation. It proposes a two-step ASH procedure that  pre-processes the standard error estimates and then follows with the ASH-t procedure on modified input \citep{lu2019empirical}.  It happens that we discover 12 peptides with differential binding affinity by two-step ASH at 10\% FDR. The different behavior of FDR-controlling procedures in the CCP+RF- RA example exposes ongoing practical challenges that are also revealed in
comprehensive numerical studies \citep{korthauer2019practical}.

Data analysts face  many issues as they filter high-dimensional measurements into short lists for experimental follow-up.  In studying this problem, we propose and evaluate a flexible empirical Bayesian mixture method that, like ASH, intervenes after effect estimates and standard errors are computed on each testing unit. The proposed \verb+MixTwice+ procedure involves shape-constrained mixture distribution for latent effects and also a separate nonparametric mixture for variance parameters (Section 2).   We leverage existing tools for constrained optimization in order to estimate the underlying mixing distributions, and we present a variety of comparative numerical experiments on the operating characteristics of \verb+MixTwice+.  The CCP+RF- RA peptide-array example happens to yield 44 peptides having significant differential antibody binding at 10\% FDR. A closer look at the identified peptides reveals binding patterns consistent with other biological information about RA, and thus provides a measure of confidence that these discoveries are not artifacts.  In a second RA example where differential signals are stronger, \verb+MixTwice+ shows a higher level of reproducibility than other approaches when presented with two independent data sets on the same populations.

\section{Mixture model}

We index peptides by $i=1, 2, \cdots, m$ and suppose that the two-group peptide-array data have been obtained and pre-processed in order to yield two summary statistics per peptide: $(x_i, s_i)$. The first component, $x_i$, is an estimated effect.  It measures the difference between the two groups, such as a difference in sample means of log-transformed data,  and is viewed a statistical estimate of an underlying effect, say $\theta_i$.  In this view, $x_i$ is a random variable having some sampling distribution, which we take to be Gaussian centered at $\theta_i$;  this is warranted noting the behavior of suitably-transformed fluorescence measurements coupled with central-limit effects for modest to large sample sizes. The second component, $s_i$, is an estimated standard error.  In the Gaussian sampling model, $\mathbb{E}(x_i)=\theta_i$ and var$(x_i)=\sigma_i^2$, and $s_i^2$ is a sample-based estimate of the variance $\sigma_i^2$.   We seek inference about the value of $\theta_i$ using local data $(x_i,s_i)$ as well as data  $\{(x_{i'},s_{i'})\}$ from  all peptides, which informs the distribution of effect and variance parameters across the array.   

Our formulation is common in large-scale inference, and we could infer $\theta_i$ values in a number of ways. For example, we could produce a peptide-specific p-value from the test statistic $t_i=x_i/s_i$ against the null 
hypothesis $H_{0,i}: \theta_i=0$.  We might refer $t_i$ to a Student-t distribution, obtain a two-sided p-value, and then process the p-values through the Benjamini-Hochberg (BH) or $q-$value methods to adjust for multiplicity \citep{benjamini1995controlling, storey2003positive}.  Alternatively, we might use the collection $\{ t_i \}$ and model their fluctuations as a discrete mixture of null and non-null cases, as in the \verb+locFDR+ procedure \citep{efron2001empirical,strimmer2008fdrtool}. Both \verb+locFDR+ and $q-$value methods are based upon discrete mixtures; interestingly the reduction of $t_i$'s to two-sided p-values entails a loss of sign information that is enough to reduce statistical power in some settings.  A more ambitious approach goes beyond null/non-null mixing to allow a full probability distribution of effects $\theta_i$ in order to account for fluctuations across all the peptides. Adaptive shrinkage (ASH) is appealing because it acquires robustness through a nonparametric treatment of this distribution, say $g(\theta)$, while using reasonable shape constraints to regularize the estimation \citep{stephens2017false}. Power advantages of ASH over other methods stem in part from its use of more data per peptide. 

In the context of an estimated mixture model there are two useful empirical-Bayesian inference statistics. The first is local false discovery rate (lfdr), $l_i=\mathbb{P}(\theta_i=0|x_i,s_i^2)$.  The term {\em local false discovery rate} was coined by Professor Efron, and the statistic may be computed in various settings beyond the specific mixture deployed in \cite{efron2001empirical}.  The list $\mathcal L$ of statistically significant peptides will be $\mathcal L= \{ i: l_i \leq c\}$ for some threshold $c$.  Notably, small $l_i$  warrants peptide $i$ to be placed in $\mathcal L$; but the value $l_i$ is also the  probability (conditional on data) that such placement is erroneous \citep{newton2006hierarchical}. Given the data, the expected rate of false discoveries in $\mathcal L$ is dominated by $c$. The local false sign rate (lfsr) is analogous to lfdr, but it avoids relying on effects being precisely zero; when the estimated effect is positive  for example, the lfsr is $\mathbb{P}(\theta_i \leq 0 | x_i, s_i^2)$. Lists controlling lfsr may be constructed in the same way as $\mathcal L$, and may be slightly smaller for the same value of $c$.  (In the CCP+RF- RA example in Section~1, ASH lfsr and lfdr lists are the same at the 10\% level.) 

With modest sample sizes, differences between estimated standard errors $\{s_i\}$ and actual standard errors $\{ \sigma_i \}$ can affect the performance of existing tools for lfdr and lfds. To better account for these differences we propose an additional mixture layer involving a sampling model $p(s_i^2|\sigma_i^2)$, which we derive from normal-theory considerations, and a flexible nonparametric mixing distribution $h(\sigma^2)$.  For both nonparametric components -- $g$ on effects $\theta_i$ and $h$  on squared standard errors $\sigma_i^2$ -- we use finite grids and treat each distribution as a vector of probabilities. We estimate $g$ and $h$ by maximum likelihood, respecting unimodal shape constraints for $g$ (as in ASH), but otherwise allowing any distributional forms.

Suppose that effects take values in a finite, regular grid $\{ a_{-K}, a_{-K+1}, \cdots, a_0, a_1, \cdots, a_K \}$ where $a_0$ is the presumed mode, taken to be $a_0=0$ in typical applications in which we aim to retain the null hypothesis of no group difference. We use $K=15$ in numerical work reported here.  Unimodality of the mixing distribution $g=(g_k)$ is
expressed as a set of ordering constraints: $g_k \geq g_{k+1}$ for $k = 0, 1, \cdots, K$ and $g_k \leq g_{k+1}$ for $k= -K, -K+1, \cdots, -1$. We also set a second regular grid $\{ 0< b_1, b_2, \cdots, b_L \}$ for squared standard errors, and impose no constraints on the mixing distribution $h=(h_l)$ aside from the basic nonparametric essentials: $h_l \geq 0$ and $\sum_l h_l=1$.

The contribution to the likelihood objective from peptide $i$ is $p(x_i,s_i^2|g,h)$:
\begin{eqnarray}
\label{unitLik}
\nonumber
 &=& \sum_k \sum_l  \mathbb{P}(\theta_i=a_k) \, \mathbb{P}(\sigma_i^2 = b_l) \, 
        p(x_i, s_i^2| \theta_i=a_k, \sigma_i^2 = b_l )   \\  \nonumber
        &=& \sum_k \sum_l g_k h_l \, p(x_i |\theta_i=a_k, \sigma_i^2=b_l ) \,
                p(s_i^2 | \sigma_i^2 = b_l ) \\
        &=& \sum_k \sum_l g_k h_l \, \frac{1}{ \sqrt{b_l}} \phi\left( \frac{x_i - a_k}{ \sqrt{b_l}}\right)
         \, \frac{\nu}{b_l} \chi_{2,\nu} \left( \frac{ \nu s_i^2 }{ b_l} \right)
\end{eqnarray}
where $\phi$ is the standard normal probability density, $\chi_{2,\nu}$ is the density of a chi-square random variable on $\nu$ degrees of freedom.  Under a normal data model, $\nu$ is determined by design (e.g. total samples minus two in the traditional two-sample comparison). The chi-square model is accurate asymptotically for a wide range of non-normal sampling distributions, however the degrees of freedom needs estimation in these cases \citep{o2014some}. 

To estimate the mixing distributions $h$ and $g$ we use the log-likelihood objective function, with terms as in~(\ref{unitLik}).
In \verb+MixTwice+, we solve the constrained optimization:
\begin{eqnarray}
\label{optimization}
\min_{g,h} -l(g,h) & = & -\sum_{i=1}^{m} \log p(x_i, s_i^2|g,h) \\
\nonumber
 {\mbox {\rm Subject to:}} & & g_k, h_l \geq 0 \quad \forall k,l \\
 \nonumber
  & & \sum_k g_k = \sum_l h_l = 1 \\
  \nonumber
  & & g_k  \leq g_{k+1}, \quad k \in \{-K, -K+1, ..., -1\} \\
  \nonumber
  & & g_k  \geq g_{k+1},  \quad k \in \{0, 1, ..., K\}
\end{eqnarray}
The gradient and Hessian of $l(g,h)$ are readily available, and so (\ref{optimization}) may be solved efficiently using augmented Lagrangian for constrained optimization, using the BFGS algorithm for inner loop optimization, which  is implemented in the R package alabama \citep{rpackagealabama}.  We extract lfdr and lfsr statistics from the peptide-specific posterior distributions at the  optimized vectors $\hat g, \hat h$: $\mathbb{P}(\theta_i = a_k|x_i, s_i^2)$

\begin{eqnarray}
\label{localpost}
\nonumber
 &=& \sum_l \mathbb{P}(\theta_i = a_k, \sigma_i^2 = b_l | x_i, s_i^2)\\ 
&\propto&  \hat g_k \sum_l  \hat h_l \, \frac{1}{ \sqrt{b_l}} \phi\left( \frac{x_i - a_k}{ \sqrt{b_l}}\right)
         \, \frac{\nu}{b_l} \chi_{2,\nu} \left( \frac{ \nu s_i^2 }{ b_l} \right).
\end{eqnarray}
Proportionality is resolved by summation over the grid $k$,
and we get:
\begin{eqnarray*}
{\rm lfdr}_i &=&  \mathbb{P}(\theta_i = a_0|x_i, s_i^2), \\ 
{\rm lfsr}_i &=& \min \left\{\sum_{k\leq 0} \mathbb{P}(\theta_i = a_k|x_i, s_i^2), \sum_{k\geq 0} \mathbb{P}(\theta_i = a_k|x_i, s_i^2) \right\}.
\end{eqnarray*}
It may be helpful to recognize that by contrast to~(\ref{localpost}), ASH-normal would entail
\begin{eqnarray}
\label{localpostASH}
 \mathbb{P}(\theta_i = a_k|x_i, s_i^2) & \propto & \hat  g_k \, \frac{1}{s_i} \phi\left( \frac{x_i - a_k}{s_i}\right),
\end{eqnarray}
and  ASH-$t$ would replace the normal density $\phi$ in~(\ref{localpostASH}) with a Student $t$ density; in both cases the ASH-estimated mixing density $\hat g$ would come not from~(\ref{optimization}) but from an objective in which mixing over variances is not explicitly accommodated.  The initial implementation of \verb+MixTwice+ invokes unimodality shape constraint, but not symmetry, and, for computational convenience, allows that a random subset of the testing units is used in the optimization.  We investigate this approximation in Supplementary Material.

\section{Simulation Study}

We are interested in the performance of \verb+MixTwice+ in scenarios reflecting what might be expected to occur in practice and have performed numerical experiments involving different generative distributions of both effects ($g$)
and variances ($h$).  Noting the special role of the null value, $\theta=0$, our experiments involve mixtures $g(\theta)= \pi_{0}\delta_0 + (1-\pi_{0})g_{{\rm alt}}(\theta)$, where $\pi_0 = \mathbb{P}(\theta_i=0)$ and $g_{\rm alt}$ provides various ways to distribute mass away from zero. Following \cite{stephens2017false} and 
\cite{lu2019empirical}, we entertain different general shapes, including so-called  {\em big-variance}, {\em bi-modal}, {\em flattop}, {\em normal}, and {\em spiky}.  \verb+MixTwice+ accounts for explicit differences between sample and underlying standard errors, and mixes nonparametrically over these underlying standard errors. Our numerical experiments consider the simplest case in which the data generating  $h$ is a point mass, a case involving a finite mixture of two values, and also a continuous case of inverse-Gamma-distributed parameters. Patterns in the error of estimation and the hypothesis testing error rates are very comparable across different choices of $h$, and so for simplicity here we report only experiments when this true $h$ is a point mass distribution at $\sigma=1$. Figures~1 and~2 summarize, respectively, properties of estimation accuracy and testing error rates.  Experiments are based on Gaussian samples, $m=1000$ peptides, and various sample size settings for the two-group comparison.  

If a method tends to overestimate $\pi_0$, then power may be reduced; in case of underestimation the FDR may be inflated.  Figure~1, Panel~B, focuses on the estimation of this marginal null frequency for one choice of sample size, namely $n=10$ observations per group. In each setting of $g$ (column), 500 data sets are generated, each drawn after its own $\pi_0$ value was uniformly drawn in $[0.5,1]$.  All methods respond appropriately to changes in $\pi_0$, though they exhibit different biases; \verb+MixTwice+ tracks the identity line (no bias) case closely in all scenarios except the challenging {\em spiky} case of $g_{\rm alt}$.  By contrast \verb+locFDR+ is conservatively biased, tending to over-estimate $\pi_0$ in most cases. Our experiments include an oracle case, namely ASH$-$normal, which takes the true value $\sigma_{i}^2=1$ as known. This numerical control helps us gauge the magnitude of statistical errors induced by  estimation error of the variance profile.

\begin{figure}
\centering
\includegraphics[width=1\columnwidth]{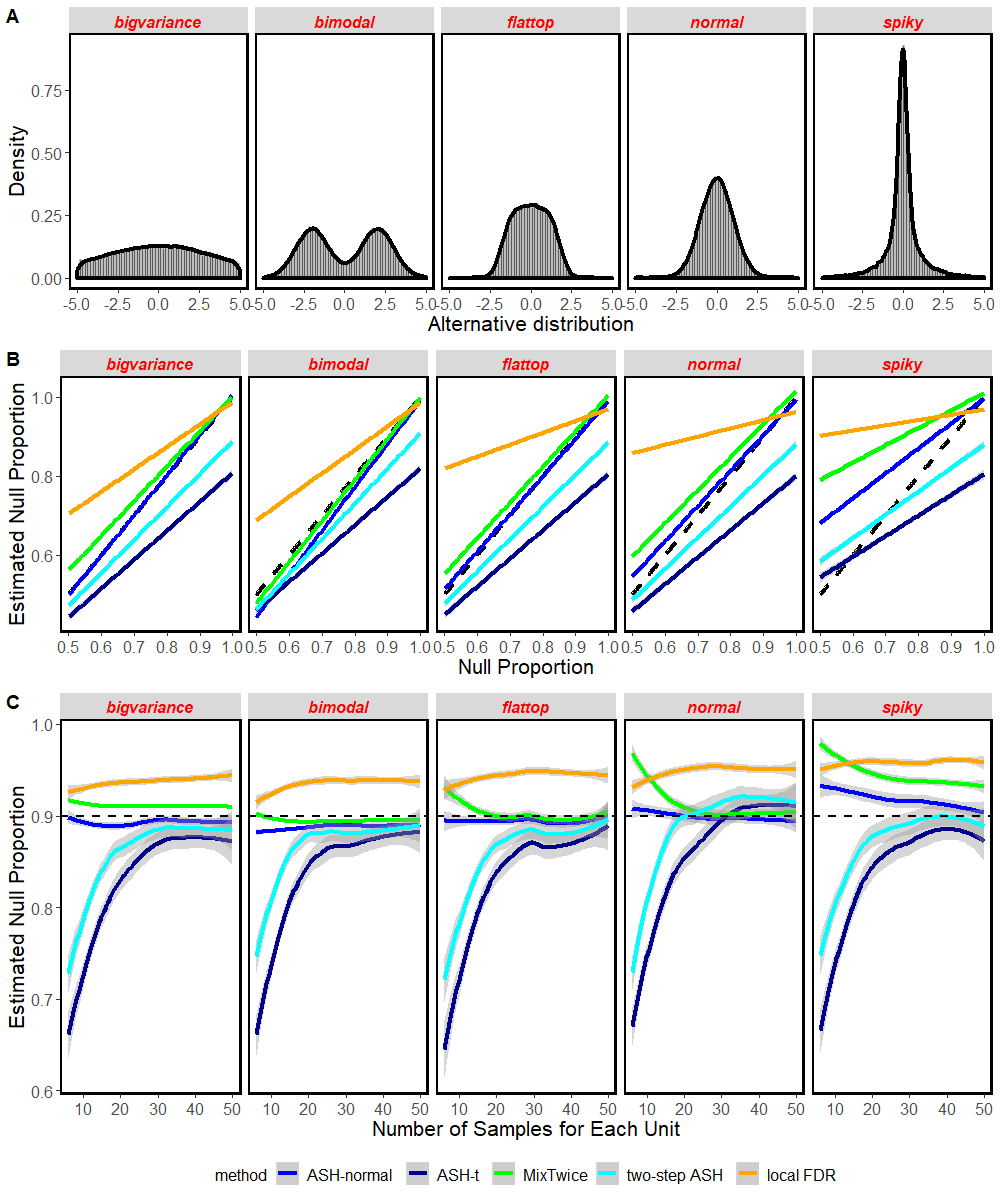}
\caption{{\bf Errors in Estimation of $\pi_0$}. Panel A shows distributions used for  $g_{{\rm alt}}(\theta)$. Panel B shows the estimation of null proportion $\pi_0$ in case of equal samples in each group of 10. Methods are distinguished by color, where we report average parameter estimates from 
500 simulated data sets.The identity line (dashed) indicates no bias. ASH-normal is an oracle case in which $\sigma_i^2=1$ is provided to the algorithm. Panel C shows error estimation as the number of observations grows, in  $\pi_0 = 0.9$. }
\label{fig:nullproportion}
\end{figure}

Figure~\ref{fig:nullproportion}, Panel~C, amplifies one case from the second row, when $\pi_0=0.9$, and shows how estimation error drops as the sample size per peptide grows. Most methods display a level of convergence in this setting, with \verb+MixTwice+ performing relatively well especially at low sample sizes.  Going beyond the estimation of $\pi_0$, we compared methods by their 1-Wassertstein error in estimating the entire mixture distribution $g$; \verb+MixTwice+ showed relatively small error in this setting also (data not shown). Without parametric assumptions, $\pi_0$ is not identifiable and only a upper bound may be reliably estimated  \citep{efron2001empirical, stephens2017false}.

\begin{figure}
\centering
\includegraphics[width=1\columnwidth]{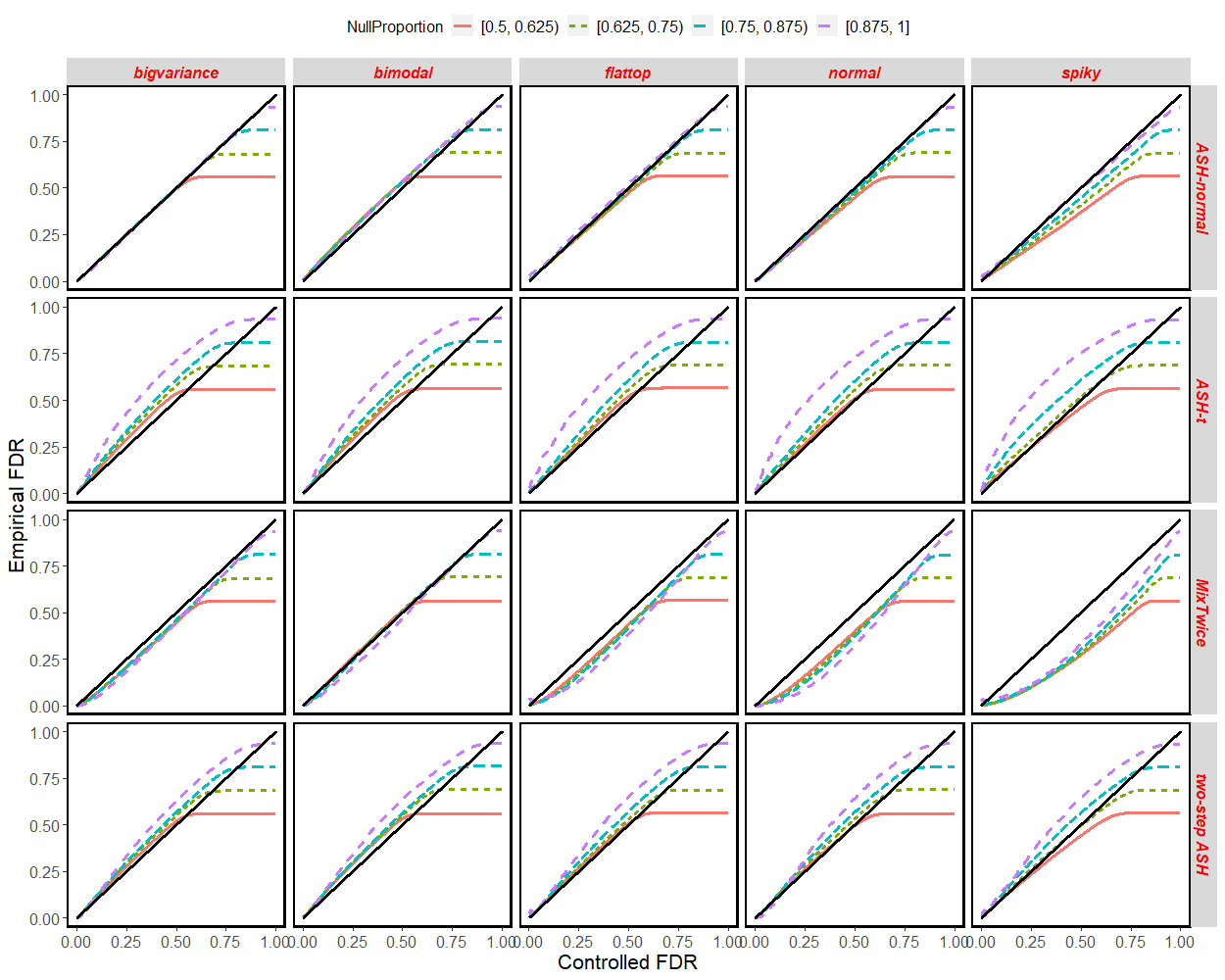}
\caption{{\bf Synthetic data and FDR control:}  False discovery
rates are shown by different methods (rows) under different  alternative distributions $g_{{\rm alt}}(\theta)$ (columns). Empirical FDR (vertical) is the
achieved error rate in the simulation;  controlled FDR (horizontal) is
the rate targeted by the methodology. Results with different $\pi_0$ are coded using different colors. A method tends to inflate FDR over the target level if
its curve is greater than the identity line; it is conservative when its 
curve is dominated by the identity line.}
\label{fig:fdr}
\end{figure}

Figure~\ref{fig:fdr} confirms that most methods are controlling FDR as advertised. The empirical false discovery rate is plotted against the controlled rate; the latter is the nominal target FDR value where we threshold the lfdr's; the former is what is evident from knowing the simulation states (in other terminology, it is the average, over simulated data sets, of the false discovery proportion).  Colored lines are used to distinguish different levels of $\pi_0$,  when the signal is dense (with a lower null proportion $\pi_0$) or when the signal is sparse (with a higher null proportion $\pi_0$). Recall we simulated independent data sets each governed by
a randomly chosen $\pi_0$ from $[0.5, 1]$.  In order to visualize the results, we stratified data sets into four groups and averaged internally: $0.5 \leq \pi_0 \leq 0.625$, $0.625 \leq \pi_0 \leq 0.75$, $0.75 \leq \pi_0 \leq 0.875$, $0.875 \leq \pi_0 \leq 1$.  The FDR inflation by ASH-t at high $\pi_0$ is evident in this simulation.

\section{Empirical studies}
\subsection{Antibodies in rheumatoid arthritis}
Rheumatoid arthritis (RA) is a chronic autoimmune disease characterized by inflammation and pain, primarily in the joints. RA patients produce autoantibodies against many different "self" proteins. Most famously, they generate antibodies against proteins in which arginine amino acids have been post-translationally modified to citrullines \citep{schellekens1998citrulline} as well as antibodies that bind to antibodies, called rheumatoid factor (RF) \citep{waaler1940occurrence}. Both autoantibody types appear to be pathogenic \citep{sokolove2014rheumatoid} and both are used diagnostically \citep{aletaha20102010}, the former detected by the anti-cyclic citrullinated peptide (CCP) test. Most RA patients make both autoantibody types (CCP+RF+ RA), but some have only one type like in CCP+RF- RA. Little is known about why CCP+RF+ versus CCP+RF- RA develops. However, a better understanding of the autoantibody repertoires in each RA subset could provide insights, a task for which peptide arrays are perfect.

The custom high-density peptide array reported in \cite{zheng2020disordered} probed 172,828 distinct 12 amino acid length peptides derived from 122 human proteins suspected to be involved in RA, including peptides in which all arginines were replaced by citrullines.  We reconsider here two distinct comparisons from that study, namely the comparison between CCP+RF- RA patients and controls, and a second comparison between CCP+RF+ RA patients and controls, in which differential signals are much stronger. Both comparisons have 12 subjects in each group. To assess reproducibility, we take advantage of a second peptide array data set derived from an independent set of 8 controls and 8 CCP+RF+ RA patients.     

\subsection{CCP+RF- RA: weak signals}\label{CCP+RF- example}
\begin{figure}
\centering
\includegraphics[width=1\columnwidth]{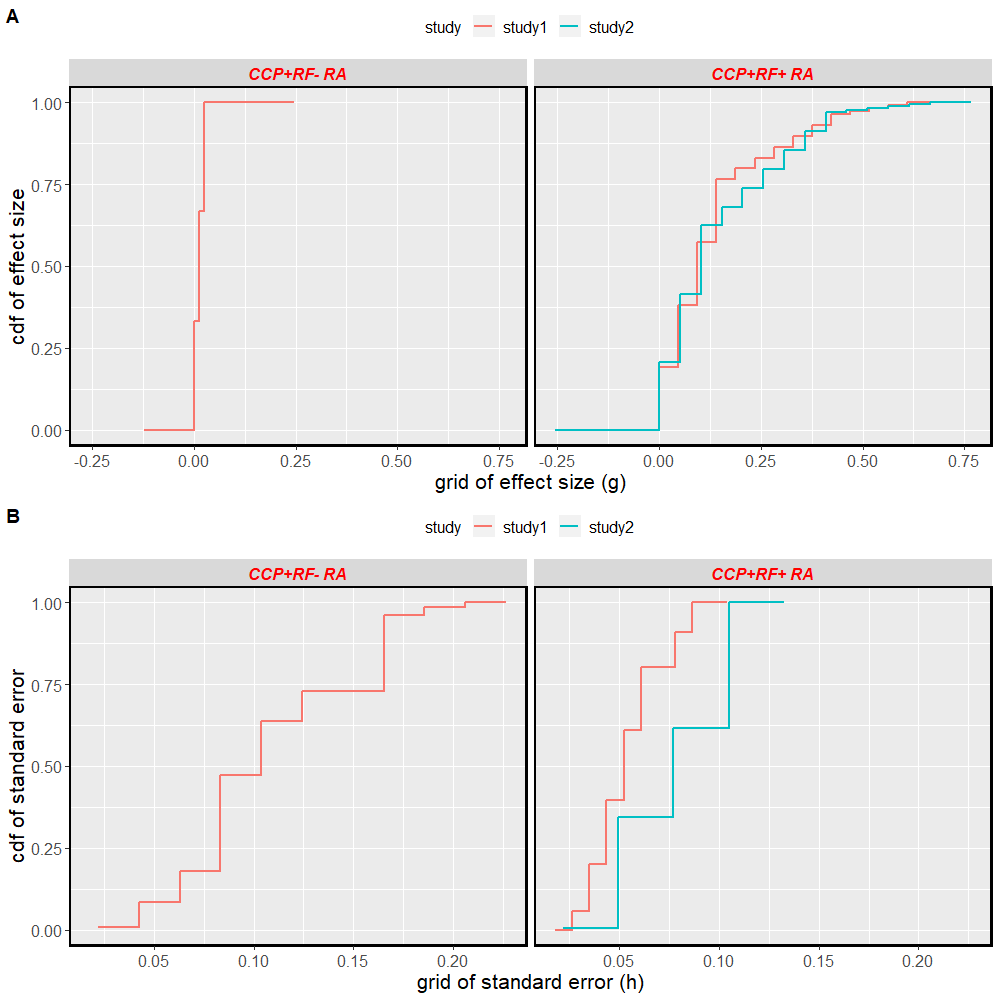}
\caption{{\bf Estimated mixing distributions:} For both effect distribution $g$ (Panel A) and squared-standard-error distribution $h$ (Panel B), shown are the maximum likelihood estimated mixing distributions as cumulative distribution functions (cdfs) in double natural log scale.  The CCP+RF- RA example is shown on the left and the two  CCP+RF+ RA examples are on the right.} 
\label{fig:mixing distribution}
\end{figure}

We applied \verb+MixTwice+ to fit the shape-constrained mixture model of Section~2. Fitted mixing distributions
are visualized in Figure~\ref{fig:mixing distribution} and provide a measure of the magnitude of changes in mean antibody levels as well as the magnitude of sampling variation.  For example, the effect-size distribution estimates no probability for effects larger than 0.037. Also, the median standard error is 0.10 (squared standard error 0.01), which is large compared to the probable effect sizes. 

\begin{figure}
\centering
\includegraphics[width=1\columnwidth]{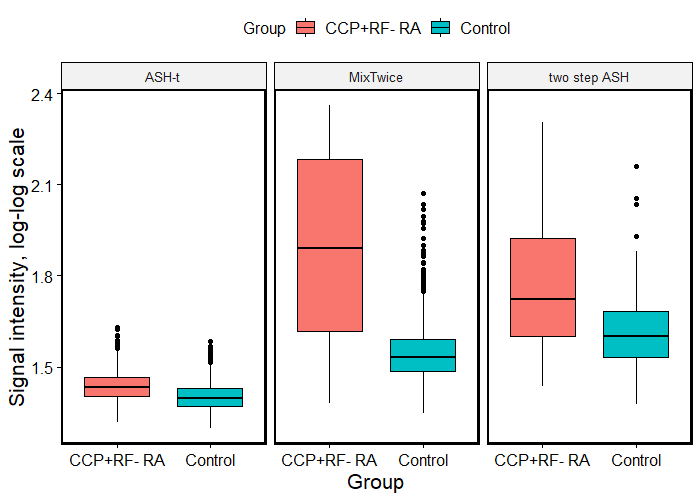}
\caption{{\bf Signal intensity of differentially abundant peptides:} Boxplots show averaged signal values on double natural log scale (both CCP+RF- RA and control subjects) for peptides found by ASH-t (76 peptides), MixTwice (44 peptides), and two-step ASH (11 peptides) all discovered at 10\% FDR.} 
\label{fig:averaged peptide boxplot for CCP+RF- group}
\end{figure}

In Section~1 we presented summary counts of peptides identified at 10\% FDR that exhibit differential binding between CCP+RF- RA patients and non-RA controls. \verb+MixTwice+, ASH-t, and two-step ASH distinguish themselves in being the only methods among many standard large-scale tools to populate non-empty lists of discovered peptides at that FDR level. Recognizing that the magnitude of signal intensities on the peptide array is an important aspect of downstream analysis, Figure~\ref{fig:averaged peptide boxplot for CCP+RF- group} shows a summary of the identified peptides by various methods. Notably, \verb+MixTwice+ and two-step ASH detect peptides in this case with higher average signal intensity than ASH-t; these may correspond to higher antibody abundance or affinity and potentially easier validation. ASH-t tends to select peptides with low standard errors, even when the estimated effects are very low. 
 
\begin{figure}
\centering
\includegraphics[width=1\columnwidth]{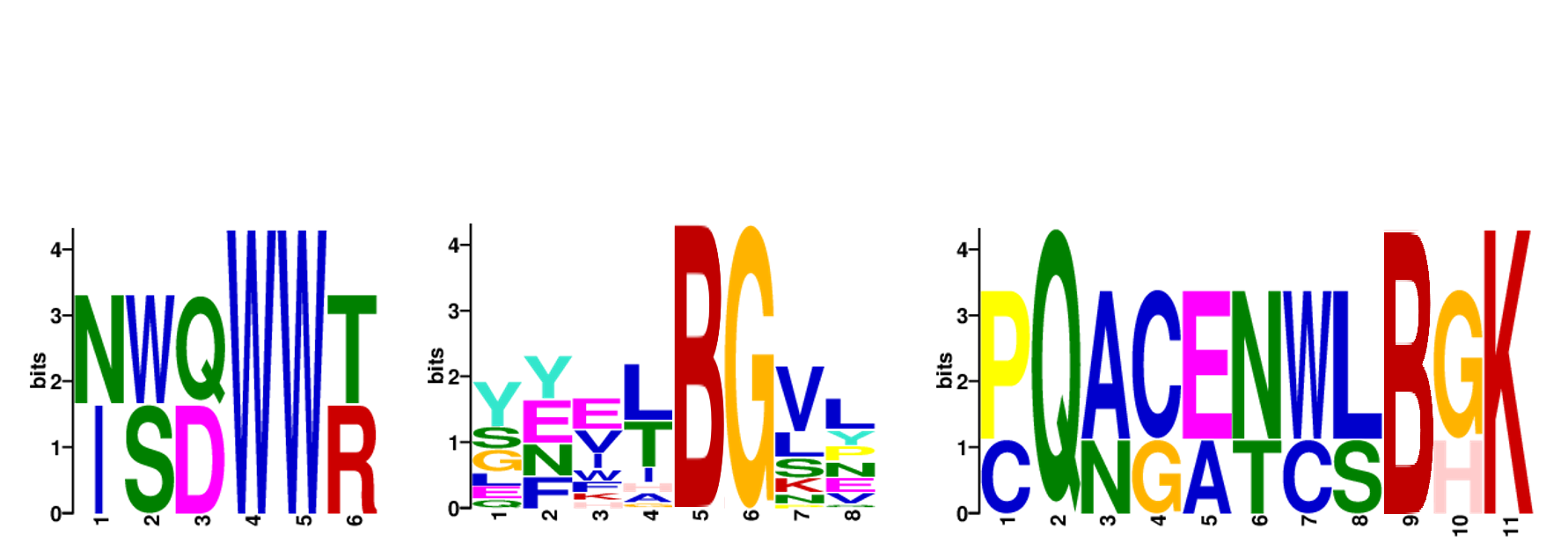}
\caption{{\bf Motif logo for significant peptides in CCP+RF- RA:} Consensus sequences were generated using online software MEME Suite \citep{bailey2009meme} and the significant peptides from the different methods: ASH-t (left), MixTwice (middle) and two-step ASH (right). Each position of the motif logo represents the empirical distribution of amino acids at that site, with size proportional to frequency. $B$ found in the middle and right panels is citrulline, a post-transitionally modified arginine. The overall height of each stack is an information measure (bits) related to the concentration of the empirical distribution on its support.}
\label{fig:motif for CCP+RF- group}
\end{figure}

Interestingly, the 44 peptides found by \verb+MixTwice+ have a strong pattern in their peptide sequences: all are citrulline ($B$)-containing peptides (which would be predicted for CCP+ RA patients) and  contain citrulline next to glycine ($B$-$G$ or $G$-$B$), as shown in the motif in Figure~\ref{fig:motif for CCP+RF- group}. Binding of antigens in which citrulline is next to glycine is consistent with a growing body of literature on the reactivity of anti-citrullinated protein antibodies in RA \citep[e.g.,][]{burkhardt2002epitope, szarka2018affinity, steen2019recognition, zheng2020disordered}.

As a further negative control calculation, we applied \verb+MixTwice+ to each of 500 permuted data sets obtained by fixing the peptide data and randomly shuffling the 24 subject labels (12 control, 12 CCP+RF- RA).  In 493 cases, the 10\% FDR list is empty; 6 cases find a single peptide and one case finds 2 peptides at this threshold. 

Among a number of large-scale testing methods applied 
to the  CCP+RF- RA example, \verb+MixTwice+ identifies a comparatively large number of statistically significant peptides. By contrast to other methods, these peptides  contain patterns in their amino acid sequences consistent with emerging evidence on this disease, and they correspond to relatively high fluorescence intensity measurements. Together, these observations provide some assurance that the \verb+MixTwice+ findings are not artifacts.

\subsection{CCP+RF+ RA: strong signals}\label{CCP+RF+ example}

One of the findings from \cite{zheng2020disordered} concerns the extensive antibody-profile differences between RA patients who are positive for both biomarkers (CCP+RF+) and control subjects.  Statistically, it represents an interesting non-sparse, large-scale testing situation, and the immunological mechanisms driving this remain only partially understood. To check the reproducibility of peptide-array findings, a new experiment was performed using the same procedures and 172,828 peptide array to detect IgG binding as in \cite{zheng2020disordered}, but with serum samples from 16 different subjects: 8 CCP+RF+ RA and 8 controls. CCP+RF+ RA and control subjects were similar in regards to age, sex, race, ethnicity, and overall health. Preprocessing followed the same protocol and provided a data set ({\em study 2}) for us to look at reproducibility of large-scale hypothesis testing methods.

\begin{figure}
\centering
\includegraphics[width=1\columnwidth]{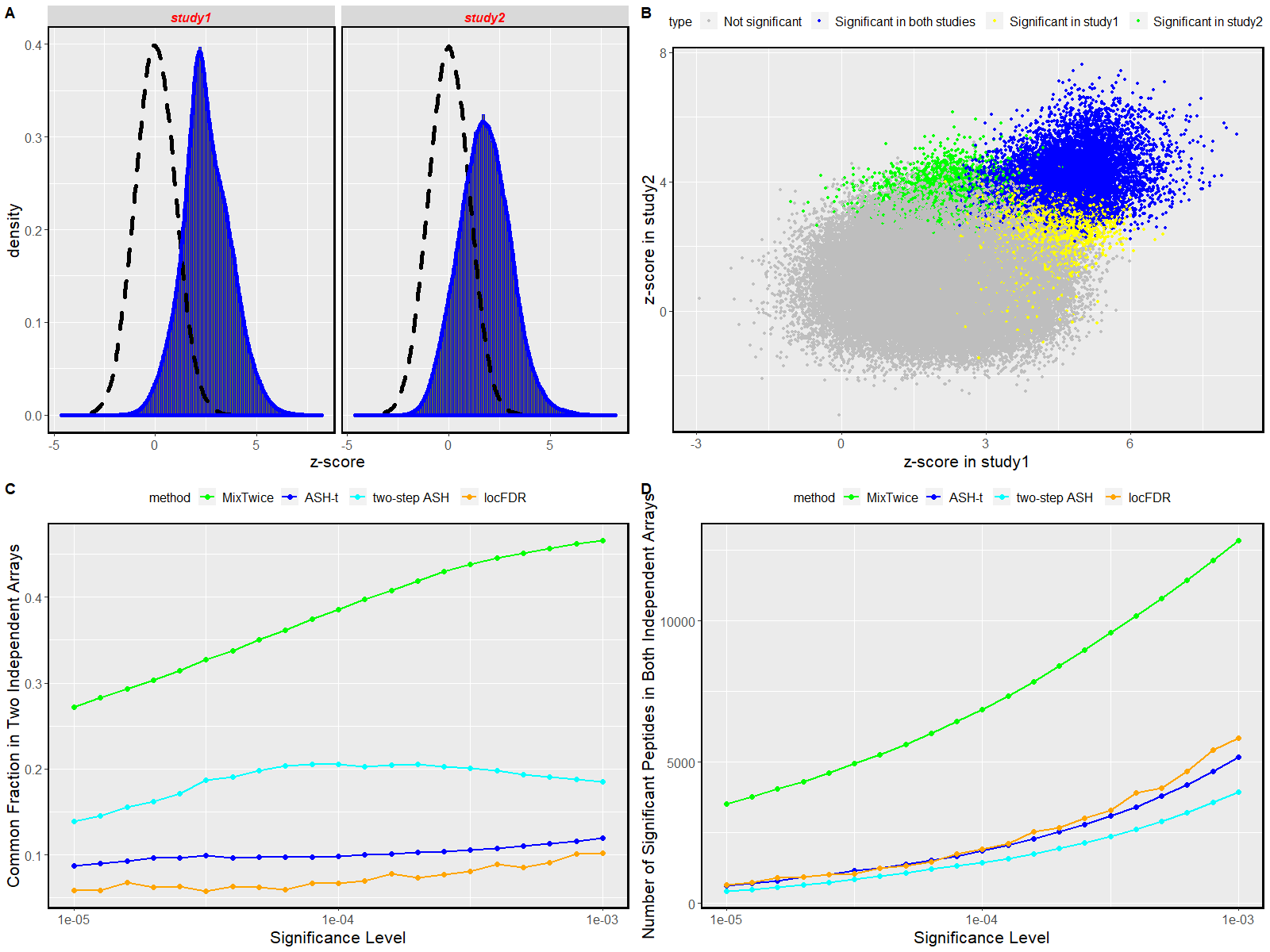}
\caption{{\bf Reproducibility comparison.} Panel A shows empirical z-score distributions for CCP+RF+ RA vs control at 172,828 peptides in two independent studies. The scatterplot in Panel B highlights peptides identified uniquely at 0.1\% FDR by MixTwice in either study (yellow, green) and those reproducibly found in both studies (blue).  Metrics in Panels~C and~D compare performance of MixTwice as a function of FDR threshold. } 
\label{fig:CCP+RF+}
\end{figure}

Z-score histograms in Figure~\ref{fig:CCP+RF+}, Panel~A, show that both studies reveal extensive increased antibody binding in the CCP+RF+ RA group. The scatterplot in Panel~B reveals concordance between the studies on this z-score metric.  The color-coding highlights discovered peptides at the 0.1\% FDR method by \verb+MixTwice+, both uniquely in one study (green or yellow) and reproducibly in both studies (blue). Of course \verb+MixTwice+ uses more information than is in the z-score summary, but the scatterplot provides a convenient visualization. The lower panels in Figure~\ref{fig:CCP+RF+} compare reproducibility statistics of different testing methods at various FDR thresholds. Denoting by $\mathcal{L}_j(\alpha)$ the list of significant peptides in study $j$ and FDR level $\alpha$, we have $|\mathcal{L}_1(\alpha) \cap \mathcal{L}_2(\alpha))|$ as the number of  peptides identified in both studies (Panel~D) and $\frac{|\mathcal{L}_1(\alpha) \cap \mathcal{L}_2(\alpha))|}{|\mathcal{L}_1(\alpha) \cup \mathcal{L}_2(\alpha))|}$ as the common fraction (Panel~C). By connecting separate, independent studies
of the same group difference, these statistics measure the reproducibility of various large-scale testing methods.  \verb+MixTwice+ shows substantially better reproducibility than other testing methods, such as ASH-t, two-step ASH, and \verb+locFDR+ in this example.
\section{Discussion}

High-throughput biomedical experiments, such as those involving peptide arrays and immunological studies, continue to provide challenging problems for large-scale hypothesis testing.  Readily applied techniques, such as $q-$value, \verb+locFDR+, \verb+IHW+, and ASH are often very effective at reporting lists of testing units (peptides) showing statistically significant effects at a targeted false discovery rate. In the case of high-density peptide arrays, we find several examples where these tools are deficient.  One issue is the number of testing units, which is an order of magnitude larger than what is seen in transcript studies, for example.  In the CCP+RF- RA comparison, most existing tools exhibit low power, which may stem in part from when they intervene in the data analysis. Methods that intervene earlier have access to more information and thereby may gain some advantage. The risk to intervening early is that more assumptions may be required to deliver relevant testing statistics (e.g., lfdr, lsdr). We rely on external validation, such as on sequence properties of the identified peptides, to assess practical utility. The CCP+RF+ RA example showcases a situation where power is high by all methods, and the differences boil down to how testing units are prioritized.  The proposed \verb+MixTwice+ procedure shows impressive reproducibility in this case.  

Structurally, \verb+MixTwice+ is similar to the ASH method for
large-scale testing: it aims to estimate a mixing distribution
of effects in an empirical Bayesian formulation.  It adopts
ASH's nonparametric, shape-constrained model for effects, but
deviates from that approach by incorporating a second mixing layer over
underlying effect-variance parameters. A number of methodological issues deserve further study. For example, \verb+MixTwice+ treats the sampling model of squared standard errors as chi-square on a design-based degrees of freedom, which is rooted in a normal-data model.  We expect that suitable transformation of the original data will make this treatment reasonable; for example, \cite{zheng2020disordered} proposed a double-log transform to stabilize variance. An interesting alternative is to use a bootstrap scheme to assess the sampling distributions directly, in order to thereby estimate the degrees of freedom that would be justified asymptotically for non-normal cases.
 
There are computational issues that warrant further investigation. The objective function~(\ref{optimization}) may not be convex in the pair of arguments $(g,h)$. Numerical experiments indicate good performance of the augmented Lagrangian optimization approach in a range of scenarios, though alternative approaches may have benefits. For example, the conditional optimizations of $g$ given $h$ or $h$ given $g$ are both convex, though attempts so far to leverage this have been less computationally efficient than the augmented Lagrangian method. Related to this are questions of grid sizes $K$ and $L$, which have to balance fidelity to the data and computational efficiency.
 
Though our presentation has focused on the classical two-group comparison problem, it should be evident that the core methodology is not restricted to this case.  Estimated effects $x_i$, for example, could arise from a contrast of interest after adjusting for blocking variables or other covariates. These will be  useful to consider as we expect them to emerge in experiments that further investigate mechanisms of immune-system disregulation.  

Finally, we point out that other forms of information may be usefully integrated with the testing methodology. Peptides tile proteins, though we have treated them as anonymous testing units. More sophisticated peptide prioritization could leverage amino-acid structure, protein content, or other features of the immunological context.

\section*{Acknowledgements}
This work was supported by the Peer Reviewed Medical Research Program (US Army Medical Research, W81XWH1810717) as well as by the University of Wisconsin-Madison, Office of the Vice Chancellor for Research and Graduate Education with funding from the Wisconsin Alumni Research Foundation to MAS and also
NIH R01 GM102756, NIH P50 DE026787, and NSF 1740707 supporting MAN. I.M.O. acknowledges support by the Clinical and Translational Science Award (CTSA) program, through the NIH National Center for Advancing Translational Sciences (NCATS), grants UL1TR002373 and KL2TR002374. This research was also supported by the Data Science Initiative grant from the University of Wisconsin-Madison Office of the Chancellor and the Vice Chancellor for Research and Graduate Education (with funding from the Wisconsin Alumni Research Foundation) (I.M.O.).  The authors acknowledge Sean McIlwain for assistance with  reproducibility calculations.

\newpage
\bibliography{references}

\begin{thebibliography}{28}
\providecommand{\natexlab}[1]{#1}
\providecommand{\url}[1]{\texttt{#1}}
\expandafter\ifx\csname urlstyle\endcsname\relax
  \providecommand{\doi}[1]{doi: #1}\else
  \providecommand{\doi}{doi: \begingroup \urlstyle{rm}\Url}\fi

\bibitem[Aletaha et~al.(2010)Aletaha, Neogi, Silman, Funovits, Felson,
  Bingham~III, Birnbaum, Burmester, Bykerk, Cohen, et~al.]{aletaha20102010}
D.~Aletaha, T.~Neogi, A.~J. Silman, J.~Funovits, D.~T. Felson, C.~O.
  Bingham~III, N.~S. Birnbaum, G.~R. Burmester, V.~P. Bykerk, M.~D. Cohen,
  et~al.
\newblock 2010 rheumatoid arthritis classification criteria: an american
  college of rheumatology/european league against rheumatism collaborative
  initiative.
\newblock \emph{Arthritis \& rheumatism}, 62\penalty0 (9):\penalty0 2569--2581,
  2010.

\bibitem[Bailey et~al.(2017)Bailey, Buechler, Matson, Peterson, Brunner, Mohns,
  Breitbach, Stewart, Ericsen, Newman, et~al.]{bailey2017pegivirus}
A.~L. Bailey, C.~R. Buechler, D.~R. Matson, E.~J. Peterson, K.~G. Brunner,
  M.~S. Mohns, M.~Breitbach, L.~M. Stewart, A.~J. Ericsen, C.~M. Newman, et~al.
\newblock Pegivirus avoids immune recognition but does not attenuate
  acute-phase disease in a macaque model of hiv infection.
\newblock \emph{PLoS pathogens}, 13\penalty0 (10):\penalty0 e1006692, 2017.

\bibitem[Bailey et~al.(2020)Bailey, Berry, Travassos, Ouattara, Boudova,
  Dotsey, Pike, Jacob, Adams, Tan, et~al.]{bailey2020microarray}
J.~A. Bailey, A.~A. Berry, M.~A. Travassos, A.~Ouattara, S.~Boudova, E.~Y.
  Dotsey, A.~Pike, C.~G. Jacob, M.~Adams, J.~C. Tan, et~al.
\newblock Microarray analyses reveal strain-specific antibody responses to
  plasmodium falciparum apical membrane antigen 1 variants following natural
  infection and vaccination.
\newblock \emph{Scientific reports}, 10\penalty0 (1):\penalty0 1--12, 2020.

\bibitem[Bailey et~al.(2009)Bailey, Boden, Buske, Frith, Grant, Clementi, Ren,
  Li, and Noble]{bailey2009meme}
T.~L. Bailey, M.~Boden, F.~A. Buske, M.~Frith, C.~E. Grant, L.~Clementi,
  J.~Ren, W.~W. Li, and W.~S. Noble.
\newblock Meme suite: tools for motif discovery and searching.
\newblock \emph{Nucleic acids research}, 37\penalty0 (suppl\_2):\penalty0
  W202--W208, 2009.

\bibitem[Benjamini and Hochberg(1995)]{benjamini1995controlling}
Y.~Benjamini and Y.~Hochberg.
\newblock Controlling the false discovery rate: a practical and powerful
  approach to multiple testing.
\newblock \emph{Journal of the Royal statistical society: series B
  (Methodological)}, 57\penalty0 (1):\penalty0 289--300, 1995.

\bibitem[Bourgon et~al.(2010)Bourgon, Gentleman, and
  Huber]{bourgon2010independent}
R.~Bourgon, R.~Gentleman, and W.~Huber.
\newblock Independent filtering increases detection power for high-throughput
  experiments.
\newblock \emph{Proceedings of the National Academy of Sciences}, 107\penalty0
  (21):\penalty0 9546--9551, 2010.

\bibitem[Burkhardt et~al.(2002)Burkhardt, Koller, Engstr{\"o}m, Nandakumar,
  Turnay, Kraetsch, Kalden, and Holmdahl]{burkhardt2002epitope}
H.~Burkhardt, T.~Koller, {\AA}.~Engstr{\"o}m, K.~S. Nandakumar, J.~Turnay,
  H.~G. Kraetsch, J.~R. Kalden, and R.~Holmdahl.
\newblock Epitope-specific recognition of type ii collagen by rheumatoid
  arthritis antibodies is shared with recognition by antibodies that are
  arthritogenic in collagen-induced arthritis in the mouse.
\newblock \emph{Arthritis \& Rheumatism}, 46\penalty0 (9):\penalty0 2339--2348,
  2002.

\bibitem[Efron(2012)]{efron2012large}
B.~Efron.
\newblock \emph{Large-scale inference: empirical Bayes methods for estimation,
  testing, and prediction}, volume~1.
\newblock Cambridge University Press, 2012.

\bibitem[Efron et~al.(2001)Efron, Tibshirani, Storey, and
  Tusher]{efron2001empirical}
B.~Efron, R.~Tibshirani, J.~D. Storey, and V.~Tusher.
\newblock Empirical bayes analysis of a microarray experiment.
\newblock \emph{Journal of the American statistical association}, 96\penalty0
  (456):\penalty0 1151--1160, 2001.

\bibitem[Hoefges et~al.(2020)Hoefges, Erbe-Gurel, McIlwain, Melby, Xu, Mathers,
  Rakhmilevich, Hank, Baniel, Pinapati, et~al.]{hoefges2020thousands}
A.~Hoefges, A.~K. Erbe-Gurel, S.~J. McIlwain, A.~S. Melby, A.~Xu, N.~Mathers,
  A.~L. Rakhmilevich, J.~A. Hank, C.~Baniel, R.~Pinapati, et~al.
\newblock Thousands of new antigens are recognized in mice via endogenous
  antibodies after being cured of a b78 melanoma via immunotherapy, 2020.

\bibitem[Ignatiadis et~al.(2016)Ignatiadis, Klaus, Zaugg, and
  Huber]{ignatiadis2016data}
N.~Ignatiadis, B.~Klaus, J.~B. Zaugg, and W.~Huber.
\newblock Data-driven hypothesis weighting increases detection power in
  genome-scale multiple testing.
\newblock \emph{Nature methods}, 13\penalty0 (7):\penalty0 577--580, 2016.

\bibitem[Korthauer et~al.(2019)Korthauer, Kimes, Duvallet, Reyes, Subramanian,
  Teng, Shukla, Alm, and Hicks]{korthauer2019practical}
K.~Korthauer, P.~K. Kimes, C.~Duvallet, A.~Reyes, A.~Subramanian, M.~Teng,
  C.~Shukla, E.~J. Alm, and S.~C. Hicks.
\newblock A practical guide to methods controlling false discoveries in
  computational biology.
\newblock \emph{Genome biology}, 20\penalty0 (1):\penalty0 1--21, 2019.

\bibitem[Lu and Stephens(2019)]{lu2019empirical}
M.~Lu and M.~Stephens.
\newblock Empirical bayes estimation of normal means, accounting for
  uncertainty in estimated standard errors.
\newblock \emph{arXiv preprint arXiv:1901.10679}, 2019.

\bibitem[Mishra et~al.(2018)Mishra, Caciula, Price, Thakkar, Ng, Chauhan, Jain,
  Che, Espinosa, Cruz, et~al.]{mishra2018diagnosis}
N.~Mishra, A.~Caciula, A.~Price, R.~Thakkar, J.~Ng, L.~V. Chauhan, K.~Jain,
  X.~Che, D.~A. Espinosa, M.~M. Cruz, et~al.
\newblock Diagnosis of zika virus infection by peptide array and enzyme-linked
  immunosorbent assay.
\newblock \emph{MBio}, 9\penalty0 (2), 2018.

\bibitem[Newton et~al.(2006)Newton, Wang, and
  Kendziorski]{newton2006hierarchical}
M.~Newton, P.~Wang, and C.~Kendziorski.
\newblock Hierarchical mixture models for expression profiles.
\newblock In \emph{Bayesian inference for gene expression and proteomics},
  chapter~2, pages 40--52. Cambridge University Press New York, 2006.

\bibitem[O’Neill(2014)]{o2014some}
B.~O’Neill.
\newblock Some useful moment results in sampling problems.
\newblock \emph{The American Statistician}, 68\penalty0 (4):\penalty0 282--296,
  2014.

\bibitem[Schellekens et~al.(1998)Schellekens, De~Jong, Van~den Hoogen, Van~de
  Putte, van Venrooij, et~al.]{schellekens1998citrulline}
G.~A. Schellekens, B.~De~Jong, F.~Van~den Hoogen, L.~Van~de Putte, W.~J. van
  Venrooij, et~al.
\newblock Citrulline is an essential constituent of antigenic determinants
  recognized by rheumatoid arthritis-specific autoantibodies.
\newblock \emph{The Journal of clinical investigation}, 101\penalty0
  (1):\penalty0 273--281, 1998.

\bibitem[Sokolove et~al.(2014)Sokolove, Johnson, Lahey, Wagner, Cheng, Thiele,
  Michaud, Sayles, Reimold, Caplan, et~al.]{sokolove2014rheumatoid}
J.~Sokolove, D.~S. Johnson, L.~J. Lahey, C.~A. Wagner, D.~Cheng, G.~M. Thiele,
  K.~Michaud, H.~Sayles, A.~M. Reimold, L.~Caplan, et~al.
\newblock Rheumatoid factor as a potentiator of anti--citrullinated protein
  antibody--mediated inflammation in rheumatoid arthritis.
\newblock \emph{Arthritis \& rheumatology}, 66\penalty0 (4):\penalty0 813--821,
  2014.

\bibitem[Steen et~al.(2019)Steen, Forsstr{\"o}m, Sahlstr{\"o}m, Odowd,
  Israelsson, Krishnamurthy, Badreh, Mathsson~Alm, Compson, Ramsk{\"o}ld,
  et~al.]{steen2019recognition}
J.~Steen, B.~Forsstr{\"o}m, P.~Sahlstr{\"o}m, V.~Odowd, L.~Israelsson,
  A.~Krishnamurthy, S.~Badreh, L.~Mathsson~Alm, J.~Compson, D.~Ramsk{\"o}ld,
  et~al.
\newblock Recognition of amino acid motifs, rather than specific proteins, by
  human plasma cell--derived monoclonal antibodies to posttranslationally
  modified proteins in rheumatoid arthritis.
\newblock \emph{Arthritis \& Rheumatology}, 71\penalty0 (2):\penalty0 196--209,
  2019.

\bibitem[Stephens(2017)]{stephens2017false}
M.~Stephens.
\newblock False discovery rates: a new deal.
\newblock \emph{Biostatistics}, 18\penalty0 (2):\penalty0 275--294, 2017.

\bibitem[Storey et~al.(2003)]{storey2003positive}
J.~D. Storey et~al.
\newblock The positive false discovery rate: a bayesian interpretation and the
  q-value.
\newblock \emph{The Annals of Statistics}, 31\penalty0 (6):\penalty0
  2013--2035, 2003.

\bibitem[Strimmer(2008)]{strimmer2008fdrtool}
K.~Strimmer.
\newblock fdrtool: a versatile r package for estimating local and tail
  area-based false discovery rates.
\newblock \emph{Bioinformatics}, 24\penalty0 (12):\penalty0 1461--1462, 2008.

\bibitem[Szarka et~al.(2018)Szarka, Aradi, Huber, Pozsgay, V{\'e}gh, Magyar,
  Gyulai, Nagy, Rojkovich, Kiss, et~al.]{szarka2018affinity}
E.~Szarka, P.~Aradi, K.~Huber, J.~Pozsgay, L.~V{\'e}gh, A.~Magyar, G.~Gyulai,
  G.~Nagy, B.~Rojkovich, {\'E}.~Kiss, et~al.
\newblock Affinity purification and comparative biosensor analysis of
  citrulline-peptide-specific antibodies in rheumatoid arthritis.
\newblock \emph{International Journal of Molecular Sciences}, 19\penalty0
  (1):\penalty0 326, 2018.

\bibitem[Tokarz et~al.(2020)Tokarz, Tagliafierro, Caciula, Mishra, Thakkar,
  Chauhan, Sameroff, Delaney, Wormser, Marques,
  et~al.]{tokarz2020identification}
R.~Tokarz, T.~Tagliafierro, A.~Caciula, N.~Mishra, R.~Thakkar, L.~V. Chauhan,
  S.~Sameroff, S.~Delaney, G.~P. Wormser, A.~Marques, et~al.
\newblock Identification of immunoreactive linear epitopes of borrelia
  miyamotoi.
\newblock \emph{Ticks and tick-borne diseases}, 11\penalty0 (1):\penalty0
  101314, 2020.

\bibitem[Varadhan(2015)]{rpackagealabama}
R.~Varadhan.
\newblock \emph{alabama: Constrained Nonlinear Optimization}, 2015.
\newblock URL \url{https://CRAN.R-project.org/package=alabama}.
\newblock R package version 2015.3-1.

\bibitem[Waaler(1940)]{waaler1940occurrence}
E.~Waaler.
\newblock On the occurrence of a factor in human serum activating the specific
  agglutination of sheep blood corpuscles.
\newblock \emph{Acta Pathologica Microbiologica Scandinavica}, 17\penalty0
  (2):\penalty0 172--188, 1940.

\bibitem[Yan et~al.(2019)Yan, Sun, Wang, Kobayashi, Ladd, Long, Lo, Patel,
  Sullivan, Albert, et~al.]{yan2019whole}
Y.~Yan, N.~Sun, H.~Wang, M.~Kobayashi, J.~J. Ladd, J.~P. Long, K.~C. Lo,
  J.~Patel, E.~Sullivan, T.~Albert, et~al.
\newblock Whole genome--derived tiled peptide arrays detect prediagnostic
  autoantibody signatures in non--small-cell lung cancer.
\newblock \emph{Cancer research}, 79\penalty0 (7):\penalty0 1549--1557, 2019.

\bibitem[Zheng et~al.(2020)Zheng, Mergaert, Fahmy, Bawadekar, Holmes, Ong,
  Bridges, Newton, and Shelef]{zheng2020disordered}
Z.~Zheng, A.~M. Mergaert, L.~M. Fahmy, M.~Bawadekar, C.~L. Holmes, I.~M. Ong,
  A.~J. Bridges, M.~A. Newton, and M.~A. Shelef.
\newblock Disordered antigens and epitope overlap between anti--citrullinated
  protein antibodies and rheumatoid factor in rheumatoid arthritis.
\newblock \emph{Arthritis \& Rheumatology}, 72\penalty0 (2):\penalty0 262--272,
  2020.

\end{thebibliography}
\clearpage

\clearpage

\clearpage

\clearpage

\clearpage

\clearpage

\clearpage
\section*{Supplementary Material}
\subsection*{Gradient and Hessian of optimization objective}

We derive the gradient and Hessian of the log-likelihood equation, $l(g,h)$. Recall the definition of $l(g,h)$:
\begin{eqnarray*}
    l(g,h) &=& \sum_{i=1}^{m} \log p(x_i, s_i^2|g,h)\\
    p(x_i, s_i^2|g,h) &=& \sum_k \sum_l g_k h_l \, \frac{1}{ \sqrt{b_l}} \phi\left( \frac{x_i - a_k}{ \sqrt{b_l}}\right)
         \, \frac{\nu}{b_l} \chi_{2,\nu} \left( \frac{ \nu s_i^2 }{ b_l} \right).
\end{eqnarray*}
To simplify notation we use $c_{i, k,l}$ to denote the prior, mixture density of sample $i$ on the grid $a_k, b_l$, and we let $d_i$ denote the observation density:
\begin{eqnarray*}
    c_{i, k,l} &:=& \frac{1}{ \sqrt{b_l}} \phi\left( \frac{x_i - a_k}{ \sqrt{b_l}}\right)
         \, \frac{\nu}{b_l} \chi_{2,\nu} \left( \frac{ \nu s_i^2 }{ b_l} \right)\\
    d_i &:=& p(x_i,s_i^2 |g,h)= \sum_k \sum_l g_k h_l \ c_{i,k,l}.
\end{eqnarray*}
Consider the parameter vector $(g,h)$  of length $(2K+1) + L$, where the first $2K+1$ components are for the effect mixing probabilities, $g = (g_k)$, and the remaining $L$ components are for variance mixing probabilities $h = (h_l)$. The quantities $c_{i,k,l}$ depend on the data, the support points but not the probabilities $g$ and $h$. 

Constraints are critical to the optimization; of course all elements of $g$ and $h$ must be positive and sum to unity. We also impose a unimodality constraint on $g$.  But in deploying the augmented Lagrangian method, these constraints act on the differentiable function $l(g,h)$, which we consider initially as varying freely over $2K+L+1$ Euclidean space. The gradient of $l(g,h)$ is a column vector of length $(2K+1) + L$ with the following format:
\begin{eqnarray*}
    \nabla l(g,h) &=& \left( \left(\frac{\partial l(g,h )}{\partial g}\right)', \left (\frac{\partial l(g,h)}{\partial h} \right)'\right )'
\end{eqnarray*}
where each component has the explicit form:
\begin{eqnarray*}
\frac{\partial l(g,h )}{\partial g_k} &=& \sum_{i=1}^{m} \frac{1}{d_i} \sum_l h_l \ c_{i,k,l} \\
    \frac{\partial l(g,h )}{\partial h_l} &=& \sum_{i=1}^{m} \frac{1}{d_i} \sum_k g_k \ c_{i,k,l}.
\end{eqnarray*}
The Hessian of $l(g,h)$ is a $(2K+1) + L$ by $(2K+1) + L$ matrix:
\begin{eqnarray*}
\nabla^2 l(g,h) = 
\begin{pmatrix}
    A & B \\
    B' & C
    \end{pmatrix}
\end{eqnarray*}
where matrix $A$ ($2K+1$ by $2K+1$) contains second derivative with respect to $g$, matrix $C$ ($L$ by $L$) contains second derivative with respect to $h$ and matrix $B$ ($2K+1$ by $L$) contains second derivative with respect to $g$ and $h$.

For entries of matrix $A$:
\begin{eqnarray*}
\frac{\partial^2 l(g,h)}{\partial g_k^2} & = & -\sum_{i=1}^{m}\frac{1}{d_i^2}\left ( \sum_l h_l \ c_{i,k,l}\right )^2\\
\frac{\partial^2 l(g,h)}{\partial g_{k_1} \partial g_{k_2}} & = & -\sum_{i=1}^{m}\frac{1}{d_i^2}\left ( \sum_l h_l \ c_{i,k_1,l}\right ) \left ( \sum_l h_l \ c_{i,k_2,l}\right ).
\end{eqnarray*}

For entries of matrix $C$:
\begin{eqnarray*}
\frac{\partial^2 l(g,h)}{\partial h_l^2} & = & -\sum_{i=1}^{m}\frac{1}{d_i^2}\left ( \sum_k g_k \ c_{i,k,l}\right )^2\\
\frac{\partial^2 l(g,h)}{\partial h_{l_1} \partial h_{l_2}} & = & -\sum_{i=1}^{m}\frac{1}{d_i^2}\left ( \sum_k g_k \ c_{i,k,l_1}\right ) \left ( \sum_k g_k \ c_{k,l_2}\right ).
\end{eqnarray*}
For entries of matrix $B$:
\begin{eqnarray*}
\frac{\partial^2 l(g,h)}{\partial g_k \partial h_l} & = & \sum_{i=1}^{m}\frac{1}{d_i^2}\left (c_{i, k,l}\ d_i - \sum_l h_l \ c_{i,k,l} \sum_k g_k \ c_{i,k,l} \right).
\end{eqnarray*}

\subsection*{Random subsampling}

The optimization
to compute $\hat g$ and $\hat h$ becomes computationally challenging
as the number of testing units increases. \verb+MixTwice+
provides an option for users to  use a randomly-selected subset of testing units to obtain the fitted distributions. 
Here we illustrate the compute-time improvements associated
with relatively little degradation in the quality of the estimates. 

We use the CCP+RF+ RA example to illustrate the random subsampling properties in terms of estimation error and computational benefit. Relative to the estimate obtained from half the units, we evaluate the discrepancy in distribution estimation and the user's CPU time (with Inter(R) Core(TM) i5-7400HQ CPU processor) when the \verb+prop+, the proportion of testing units used to fit the distribution, changes. We use $1$-Wasserstein distance between two cumulative distribution functions as the metric to evaluate the discrepancy from the case when  $\verb+prop+ = 0.5$ as benchmark.

\begin{figure} 
    \centering
    \includegraphics[width=1\columnwidth]{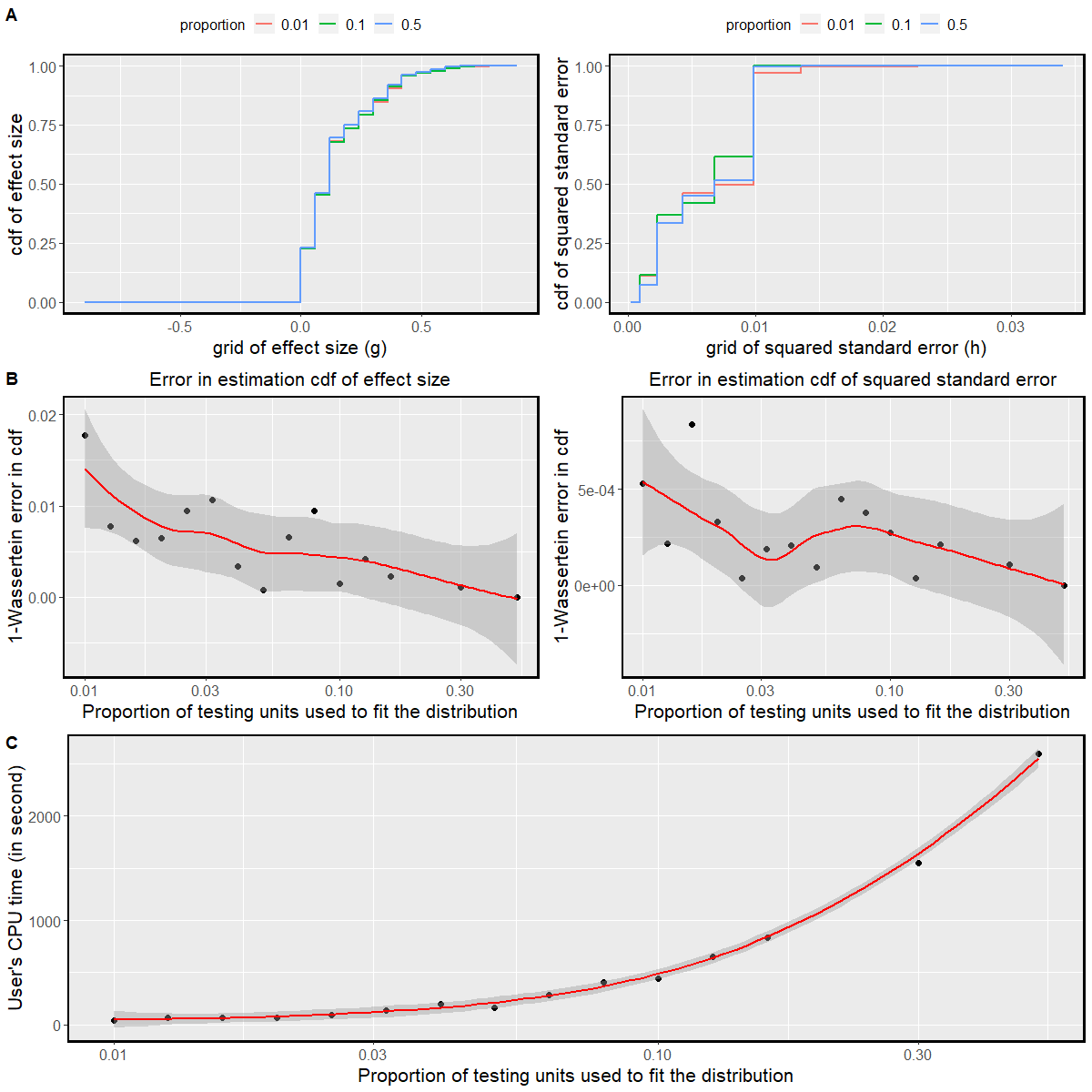}
    \caption{\textbf{How does random subsampling influence estimation accuracy and computational efficiency?} Panel~A shows the estimation in $\hat g, \hat h$ when various
    proportions of the units are used for estimation.  Panel~B shows the 1-Wasserstein discrepancy (between estimate
    at that proportion and estimate from half the units) as a function of subsampling
    proportion. Panel~C shows the corresponding CPU time.}
    \label{random subsampling}
\end{figure}

Figure~\ref{random subsampling} summaries the result. Panel~A highlights the estimation of $\hat g, \hat h$ when $\verb+prop+ = 0.5,0.1,0.01$ where the estimations are quite similar. Panel~B shows how the discrepancy decreases when the proportion of testing units used to fit the distribution increases. Note that even when $\verb+prop+ = 0.01$, the discrepancy is quite small (error in $\hat g$ less than 0.02 and error in $\hat h$ only $10^-4$). Panel~C shows the computational benefits.

\end{document}